\documentclass[12pt,a4paper]{article}
\usepackage{latexsym}
\usepackage{amssymb}
\usepackage{amsmath}
\usepackage{amsfonts}
\usepackage{mathrsfs}
\usepackage{cite}

\catcode`@=11
\renewcommand{\section}{\@startsection{section}{1}{0pt}{\medskipamount}
{\medskipamount}{\large\bf}} \numberwithin{equation}{section}
\catcode`@=12

\newcommand{\be}{\begin{equation}}
\newcommand{\ee}{\end{equation}}
\newcommand{\p}[1]{(\ref{#1})}

\def\a{\alpha}

\def\b{\beta}
\def\p{\partial}

\def\t{\theta}

\def\Tr{{\rm Tr}\,}
\def\cN{{\cal N}}

\def\cA{{\cal A}}

\def\bea{\begin{eqnarray}}
\def\eea{\end{eqnarray}}
\def\nn{\nonumber}

\def\cN{{\cal N}}

\def\f{\frac}

\def\nn{\nonumber}

\def\d{\delta}

\def\g{\gamma}

\def\ve{\varepsilon}

\def\sB{\stackrel{\frown}{\square}}

\def\eq{\eqref}

\def\nb{\nabla}

\numberwithin{equation}{section}

\date{\it  }
\begin{document}
\begin{titlepage}

\begin{center}
\vspace{1cm} {\Large\bf One-loop divergences in the $6D$, $\cN=
(1,0)$ abelian} \vspace{0.1cm}

{\Large\bf gauge theory \vspace{1.2cm} }

 {\bf
 I.L. Buchbinder\footnote{joseph@tspu.edu.ru }$^{\,a,b}$,
 E.A. Ivanov\footnote{eivanov@theor.jinr.ru}$^{\,c}$
 B.S. Merzlikin\footnote{merzlikin@tspu.edu.ru}$^{\,a,d}$,
 K.V. Stepanyantz\footnote{stepan@m9com.ru}$^{\,e}$
 }

 {\it $^a$ Department of Theoretical Physics, Tomsk State Pedagogical
 University,\\ 634061, Tomsk,  Russia \\ \vskip 0.15cm
 $^b$ National Research Tomsk State University, 634050, Tomsk, Russia \\ \vskip 0.1cm
 $^c$ Bogoliubov Laboratory of Theoretical Physics, JINR, 141980 Dubna, Moscow region,
 Russia \\ \vskip 0.1cm
 $^d$ Department of Higher Mathematics and Mathematical Physics,\\
\it Tomsk Polytechnic University, 634050, Tomsk, Russia\\
\vskip 0.1cm
 $^e$ Department of Theoretical Physics, Moscow State University,
119991, Moscow, Russia

}
\end{center}

\begin{abstract}
We consider, in the harmonic superspace approach, the
six-dimensional $\cN=(1,0)$  supersymmetric model of abelian gauge
multiplet coupled to a hypermultiplet. The superficial degree of
divergence is evaluated and the structure of possible one-loop
divergences is analyzed. Using the superfield proper-time and
background-field technique, we compute the divergent part of the
one-loop effective action depending on both the gauge multiplet and
the hypermultiplet. The corresponding counterterms contain the
purely gauge multiplet contribution together with the mixed
contributions of the gauge multiplet and hypermultiplet. We show
that the theory is on-shell one-loop finite in the gauge multiplet
sector in agreement with the results of \cite{HS}. The divergences
in the mixed sector cannot be eliminated by any field redefinition,
implying the theory to be UV divergent at one loop.

\end{abstract}
\end{titlepage}

\setcounter{footnote}{0}

\setcounter{page}{1}

\section{Introduction}
The higher-dimensional supersymmetric gauge models are of interest mainly because they describe
low-energy limits of the superstring/brane theory and inherit many remarkable
properties of the latter. In particular, one can expect the existence of various
non-renormalization theorems governing their ultraviolet (UV) behavior.
In this letter we study the UV divergence structure of the
six-dimensional abelian ${\cal N}=(1,0)$ gauge theory interacting
with hypermultiplets. The analysis of this simplest model can be
conducive for the further study of quantum properties of more
complicated non-abelian ${\cal N}=(1,0)$ and ${\cal N}=(1,1)$ gauge
theories.

An analysis of the UV divergences in the higher-dimensional
supersymmetric gauge theories has been initiated by the paper
\cite{HS} and continued in the subsequent papers \cite{HS1}, \cite{BHS},
\cite{BHS1}, \cite{ISZ}, \cite{BP15}, \cite{BP15-1}, \cite{BIS},
\cite{S} (and references therein). In particular, it was found
that in the sector of gauge (or vector) multiplet the divergences at
different loops reveal a universal structure and in many cases
some counterterms can be completely eliminated by the field
redefinitions. The counterterms in the hypermultiplet sector have never
been calculated.

As is well known, the most efficient way to describe the quantum
aspects of supersymmetric theories is to use the off-shell superfield
formulations (see e.g. \cite{BK} for $4D,\, {\cal
N}=1$ theories and \cite{GIOS} for $4D, \,{\cal N}=2$ theories). An
arbitrary $(n,m)$ representation of the six-dimensional
supersymmetry is labeled by the numbers of left ($n$) and right
($m$) independent supersymmetries (see, e.g., \cite{HST}). In the case
of $6D, \,{\cal N} = (1,0)$ supersymmetry, the models of vector multiplet and
hypermultiplet can be formulated off shell in terms of superfields defined on $6D, \,{\cal N}=(1,0)$
harmonic superspace \cite{HSW}, \cite{Z} (see also \cite{ISZ},
\cite{BP15}, \cite{BIS}  and references therein). It allows to
formulate an arbitrary six-dimensional ${\cal N}=(1,0)$ supersymmetric
Yang-Mills theory in $6D, \, {\cal N}=(1,0)$ superspace as a theory of
interacting  unconstrained off-shell superfields describing the six-dimensional
${\cal N}=(1,0)$ vector multiplet and hypermultiplet. Using the
appropriate set of ${\cal N}=(1,0)$ harmonic superfields, one can construct
${\cal N}=(1,1)$ supersymmetric Yang-Mills theories (see e.g. \cite{BIS}),
as well as the free gauge models with ${\cal N}=(2,0)$ supersymmetry \cite{BP15}. It is
worth pointing out that, in many aspects, $6D, \, {\cal N}=(1,0)$ SYM theory is analogous to $4D, \,{\cal N}=2$ SYM theory, and $6D,\, {\cal N}=(1,1)$ SYM
theory to $4D,\, {\cal N}=4$ SYM theory. These $6D$ theories and their $4D$ counterparts have equal numbers of supercharges,
8 and 16, respectively. Like $4D, \,{\cal N}=4$ SYM theory
possesses manifest off-shell ${\cal N}=2$ supersymmetry and an on-shell hidden
${\cal N}=2$ supersymmetry, $6D,\, {\cal N}=(1,1)$ SYM theory possesses manifest ${\cal N}=(1,0)$ supersymmetry and an on-shell hidden ${\cal N}=(0,1)$
supersymmetry (see \cite{BIS} for details and further references).

The general analysis of the possible low-energy contributions of
different conformal dimensions to the effective action of ${\cal
N}=(1,0)$ SYM theories has been carried out in ref. \cite{BIS}. It
was proved that the superfield counterterm of dimension 6 in ${\cal
N}=(1,0)$ harmonic superspace is a linear combination of three
terms, where one depends only on the vector multiplet superfield,
another depends only on the hypermultiplet and the third mixed term
bears a dependence on both the vector multiplet and the
hypermultiplet. The analysis was based on the ${\cal N}=(1,0)$
harmonic superspace formulation of the theory and transformation
properties of the involved ${\cal N}=(1,0)$ harmonic superfields.
Taking into account the results obtained in \cite{BIS}, it would be
extremely interesting to demonstrate how these results can in
principle be derived in the framework of quantum field theory.
Namely this problem is solved in this letter for an abelian ${\cal
N}=(1,0)$ supersymmetric gauge theory, which is an abelian ${\cal
N}=(1,0)$ vector models coupled to ${\cal N}=(1,0)$ hypermultiplet.

The paper is organized as follows. In section 2 we briefly describe
the formulation of abelian $6D, \,{\cal N}=(1,0)$ gauge theory in
${\cal N}=(1,0)$ harmonic superspace and fix the $6D$ notations and
conventions.
Section 3 presents the harmonic superspace background field method which allows one to obtain
the effective action in a manifestly gauge invariant and ${\cal N}=(1,0)$
supersymmetric form. In section 4 we derive the superficial degree
of divergence in the theory of interacting vector and hyper
multiplets and discuss the structure of the one-loop divergences. In
particular, we prove that the one-loop counterterms indeed match
with the results of \cite{BIS}, except that the purely hypermultiplet divergent
contribution to the effective action is absent in the one-loop
approximation. Section 5 is devoted to the direct calculations of the
one-loop divergences. In section 6 we summarize the results and
discuss the  problems for further study.

\section{Abelian gauge theory in $6D$, $\cN=(1,0)$ harmonic superspace}
Our consideration in this section (including notations, conventions and terminology) will closely follow ref. \cite{BIS}.

The basic objects of the $6D, {\cal N}=(1,0)$ superfield gauge
theory are gauge covariant derivatives defined by \be \nb_{\cal
M}=D_{\cal M}+i\cA_{\cal M}, \ee where $D_{\cal M} = (D_M, D^i_a)$
are the flat derivatives. Here  $M=0,..,5,$ is the $6D$ vector index
and $a=1,..4,$ is the spinorial one. The superfield $\cA_{\cal M}$
is the gauge super-connection. The covariant derivatives transform
under the gauge group as \be
 \nb'_{\cal M}=e^{{\rm i}\tau}\nb_{\cal M} e^{-{\rm i}\tau}\,, \quad
 \tau^{\dag}=\tau\,,
\label{tau}
 \ee
 and satisfy the algebra
 \bea
&& \{\nb_a^i,\nb^j_b\}=-2i\ve^{ij}\nb_{\a\b},\quad
[\nb_c^i,\nb_{ab}]=-\f12\ve_{abcd}W^{i\,d}, \label{alg1}\\
&&[\nb_M, \nb_N] = i F_{MN}\,, \label{alg2}
 \eea
where $W^{i\,a}$ is the superfield strength and $\nb_{ab} =
\f12(\g^M)_{ab} \nb_M$. Further in this paper we consider only the
abelian gauge theory coupled to a hypermultiplet.

The constraints \eqref{alg1} and \eqref{alg2} can be solved in the harmonic superspace framework.
In the $\lambda$-frame
\cite{GIOS}, the spinor covariant derivatives $\nb^+_a$ coincide
with the flat ones $D^+_a\,$, while the harmonic covariant derivatives
acquire the connections $V^{++}$ and $V^{--}$,
 \bea
 &&\nb^{\pm\pm}=D^{\pm\pm}+iV^{\pm\pm}~, \quad \widetilde{V}^{\pm\pm} = V^{\pm\pm}\,, \quad \delta V^{\pm\pm} = -\nb^{\pm\pm}\lambda(\zeta, u)\,,\label{cD++}\\
 && [\nabla^{--}, D^+_a] = \nabla^{-}_a\,, \quad [\nabla^{++}, \nabla^{-}_a] = D^+_a\,, \quad [\nabla^{++}, D^{+}_a]
= [\nabla^{--}, \nabla^{-}_a] = 0\,, \label{constrAnal}
 \eea
where $(\zeta, u)$ stands for the analytic subspace coordinates. The
real connection $V^{++}(\zeta, u)$ is analytic (in virtue of the
third constraint in \eq{constrAnal}) and it is an unconstrained
potential of the theory. The component expansion of $V^{++}(\zeta,
u)$ in the Wess-Zumino gauge reads
 \be
V^{++}_{WZ}=\t^{+ a}\t^{+
b}A_{ab}(x_{(an)})+(\t^+)^3_a\lambda^{ia}(x_{(an)})u^-_i +3(\t^+)^4D^{ik}(x_{(an)})u^-_iu^-_k\,.
 \ee
It involves  the  gauge field $A_{ab}$, the gaugino field $\lambda^{i\, a}$ and
the auxiliary field $D^{(ik)}$.

The second, non-analytic harmonic connection $V^{--}(z, u)$ is
uniquely determined in terms of $V^{++}$  as a solution of the
harmonic zero-curvature condition \cite{GIOS}. In the abelian case
the latter is
 \be
 \label{zeroc}
[\nabla^{++}, \nabla^{--}] = D^0 \; \Leftrightarrow \;  D^{++}V^{--}-D^{--}V^{++}=0.
 \ee
The equation \eq{zeroc} can be solved for $V^{--}$ as
 \be
V^{--}(z,u)=  \int du_1 \frac{V^{++}(z,u_1)}{(u^+ u^+_1)^2}\,.
 \ee
Using the connection $V^{--}$, we can construct the spinor and
vector superfield connections  and define the covariant spinor superfield
strengths $W^{\pm a}$
 \bea
W^{+a}=-\frac{1}{6}\varepsilon^{abcd}D^+_b D^+_c D^+_d V^{--}\,,
\quad W^{-a} = D^{--}W^{+a}\,.\label{W+Def}
 \eea
We also  define the Grassmann-analytic superfield \cite{BIS}
 \bea
F^{++} = \frac14 D^+_a W^{+ a} =(D^+)^4 V^{--}\,, \quad D^+_a W^{+ b} = \delta^b_a F^{++}\,,
\quad D^{++}
F^{++}=0\,, \label{identity0}
 \eea
which will be used for construction of the counterterms.

The superfield action of $6D,\,\cN=(1,0)$ abelian model of gauge
multiplet interacting with hypermultiplet  has the form \be
 \label{S0}S[V^{++}, q^+]=\f{1}{4 f^2}\int
 d^{14}z\, \f{du_1 du_2}{(u^+_1u^+_2)^2 }\, V^{++}(z,u_1)V^{++}(z,u_2)
 - \int d\zeta^{(-4)} du\, \tilde{q}^+ \nabla^{++} q^+,
 \ee
where $f$ is a dimensionful coupling constant ($[f]=-1$). The
hypemultiplet superfield $q^+(x,\t)$ has a short expansion
$q^+(z)=f^i(x)u^+_i +\t^{ +\,a}\psi_a(x) +\ldots$, with a doublet of
massless scalars fields $f^i(x)$ and the spinor field $\psi_\a$ as
the physical fields. It also involves an infinite tail of auxiliary
fields coming from the harmonic expansions. Both the superfield
$q^+(\zeta,u)$ and its $\;\widetilde{\,}\;$-conjugate $\tilde{q}^+$
\cite{GIOS} obey the analyticity constraint, $D^+_\a q^+ = D^+_\a
\tilde{q}^+=0\,$. The action \eq{S0} is invariant under the gauge
transformation
 \be
 V^{++}{}' = -i e^{i\lambda}D^{++} e^{-i\lambda}
 +e^{i\lambda} V^{++} e^{-i\lambda}\,, \qquad  q^+{}' =  e^{i\lambda}
 q^+\,,
 \label{gtr}
 \ee
where $\lambda = \lambda(\zeta,u)$ is the analytic gauge parameter introduced in \eq{cD++}.
Using the zero curvature condition \eq{zeroc}, one can derive a useful
relation between arbitrary variations of harmonic connections
\cite{BIS}
 \be
\d V^{--} = \f12(D^{--})^2\d V^{++} - \f12 D^{++}(D^{--}\delta
V^{--}) \,. \label{var}
 \ee

Classical equations of motion following from the action \eq{S0} read
 \be
 \frac{{\delta}S}{{\delta}V^{++}}=0 \,\Rightarrow \, \frac{1}{2f^2} F^{++} - i\tilde{q}^+q^+ = 0\,, \qquad
 \frac{{\delta}S}{{\delta}\tilde{q}^{++}} = 0 \,\Rightarrow \,\nabla^{++}q^+ = 0\,. \label{eqm}
  \ee
Note that the superfield $F^{++}$ is real under the $\;\widetilde{\,}\;$ conjugation, $\widetilde{F}^{++} = F^{++}$. The $\;\widetilde{\,}\;$ -
reality of the first equation in \eq{eqm} (as well as  of the action \eq{S0}) is guaranteed by the conjugation rule
$\widetilde{{\tilde q}^{+}} = - q^+$ \cite{GIOS}.

\section{Background field method}
In this section we outline the background field method for the model
\eqref{S0}. The construction of gauge invariant effective action in
the model under consideration is very similar to that for $4D,
\cN=2$ supersymmetric gauge theories \cite{BUCH1}, \cite{BP07} (see
also the reviews \cite{BUCH2})\footnote{There are two approaches for
constructing the background field method for $4D, {\cal N} = 2$ SYM
theories. One is formulated in the conventional ${\cal N} = 2$
superspace \cite{HST-1}, while another in $4D, {\cal N} = 2$
harmonic superspace \cite{BUCH1}, \cite{BKO97} (see also the reviews
\cite{BUCH2}).  The first formulation faces some troubles basically
related to an infinite number of FP ghosts. The second approach is
free of such difficulties and provides a convenient tool for
manifestly covariant loop calculations. In this paper we generalize
it to $6D, {\cal N}= (1, 0)$ gauge  theory. Though in our case the
problem of ghosts is absent at all because we deal with the abelian
theory, the harmonic background field method looks most preferable
like in $4D$ case.}.

We split the superfields
$V^{++}, q^{+}$ into the sum of the ``background'' superfields
$V^{++}, Q^{+}$ and the ``quantum'' ones $v^{++}, q^{+}\,$,
 \be
 V^{++}\to V^{++} + fv^{++}, \qquad q^{+} \to Q^{+} + q^{+}\,,
 \ee
and then expand the action in a power series in quantum fields.
As a result, we obtain the classical action as a functional of
background superfields and quantum superfields.

To construct the gauge invariant effective action, we need to impose the
gauge-fixing conditions only on quantum superfields. As in
the four-dimensional case \cite{BUCH1}, we introduce the gauge-fixing
function in the form
 \be
{\cal F}^{(+4)}=D^{++}v^{++}. \label{gf}
 \ee
We consider the abelian gauge theory, where gauge-fixing function
\eqref{gf} is background-field independent. This means that the
Faddeev-Popov ghosts are also background-field independent and so
make no contribution to the effective action. According to \eq{gf},
the gauge-fixing part of the quantum field action has the form
  \be
S_{GF} = -\frac{1}{4}\int d^{14}z du_1
du_2\frac{v^{++}(1)v^{++}(2)}{(u^+_1u^+_2)^2}+\frac{1}{8}\int
d^{14}z du v^{++} (D^{--})^2 v^{++}~. \label{SGF}
 \ee
A formal expression of the effective action $\Gamma[V^{++}, \; Q^+]$
for the theory under consideration is constructed by the
Faddeev-Popov procedure (see the reviews \cite{BUCH2} for details).

In the one-loop approximation, the first quantum correction to the classical
action $\Gamma^{(1)}[V^{++}, \; Q^+]$ is given by the following path
integral \cite{BUCH1,BP07}:
 \be
 e^{ i\Gamma^{(1)}[V^{++}, Q^+]} =
 \int{\cal D}v^{++}{\cal D}q^+ {\cal D}\tilde{q}^+\, e^{iS_{2}[v^{++},\,q^+;\,V^{++},\,
 Q^+]}\,.
 \label{Gamma0}
 \ee
Here, the full quadratic action $S_2$ is the sum of the
classical action \eq{S0}, with the  background-quantum splitting accomplished, and the
gauge-fixing action \eqref{SGF}
 \bea
 S_2 &=&\f14\int d\zeta^{(-4)}du\,
v^{++} \square_{(2,2)}v^{++}  \nn\\
&& - \int d\zeta^{(-4)}du\big\{\tilde{q}^+{\nabla}^{++}q^{+} +
f\tilde{Q}^+ iv^{++}q^{+} + f\tilde{q}^+ iv^{++} Q^{+}\big\}\,,
\label{S2}
 \eea
where the operator $\square_{(2,2)} = \f12 (D^+)^4(D^{--})^2$
transforms the analytic superfields $v^{++}$ into analytic
superfields. The Green function, associated with
$\square_{(2,2)}\,$, $G^{(2,2)}(z_1,u_1|z_2,u_2) = i\langle
v^{++}(z_1,u_1)v^{++}(z_2,u_2)\rangle\,,$ is given by the expression
similar to that in the $4D, {\cal N}=2$ case \cite{GIOS}
 \bea
G^{(2,2)}_\tau(1|2) = -2 \f{(D^+_1)^4}{\square_1} \d^{14}(z_1-z_2)
\d^{(-2,2)}(u_1,u_2)\,. \label{VGreen}
 \eea

The action $S_2$ (\ref{S2}) is a quadratic form of quantum fields,
with the coefficients depending on background fields. For further
use, it is convenient to diagonalize this quadratic form, that is to
decouple the quantum superfields $v^{++}$ and $q^{+}$. To achieve this, one
performs the special change of the quantum hypermultiplet variables
\footnote{A similar substitution was used in \cite{BP07},
\cite{Kuz07} and \cite{BM15} for computing one- and two-loop
effective actions in supersymmetric theories,  and in \cite{OV} for non-local change of fields in non-supersymmetric QED.}
in the path integral, such that it removes the mixed terms,
 \bea
 \label{replac}
 q^{+}_1= h^{+}_1 - f\int d \zeta^{(-4)}_2 du_2\, G^{ (1,1)}(1|2)\,i v^{++}_2 Q^{+}_2\,,
 \eea
with $h^{+}$ being the new independent quantum superfield. It is evident that the Jacobian of the variable change (\ref{replac}) is unity.
Here
$G^{(1,1)}(\zeta_1,u_1|\zeta_2,u_2) =
i\langle\tilde{q}^{+}(\zeta_1,u_1){q}^{+}(\zeta_2,u_2)\rangle$ is
the superfield hypermultiplet Green function in the $\tau$-frame
($G^{(1,1)}(1|2) = - G^{(1,1)}(2|1)$). This Green function is
analytic with respect to both arguments and satisfies the equation
 \bea
 \label{eqG}
 \nb_1^{++}G_\tau^{(1,1)}(1|2)=\d_A^{(3,1)}(1|2) \; \Rightarrow \; G_\tau^{(1,1)}(1|2)=
 \f{(\nb^+_1)^4(\nb^+_2)^4}{\sB_1}\f{\d^{14}(z_1-z_2)}{(u^+_1u^+_2)^3}\,,
 \label{GREEN}
  \eea
where $\d_A^{(3,1)}(1|2)$ is the covariantly-analytic delta-function
and $\sB$ is the covariantly-analytic d'Alembertian  \cite{BP15}
which acts on analytic superfields as follows
 \bea
 \sB = \f12(D^+)^4(\nb^{--})^2 = \square+ i{W}^{+\,a}\nb^-_a
+ iF^{++}\nb^{--}  - \f{i}4 (D^-_a {W}^{+a})\,, \label{Box1}
 \eea
with $\square = \f12 \varepsilon^{abcd}\nb_{ab}\nb_{cd} = \nb^M
\nb_M$. Note that the covariant d'Alembertian transforms the
analytic superfields into analytic superfields. After some algebra,
the quadratic part of the action $S_2$ \eq{S2} splits into the
vector-multiplet dependent part
 \bea
 S_{2}^{Vect}[V^{++}, Q^+] &=& \f14\int d\zeta^{(-4)}_1
 du_1\,v^{++}_1 \nn \\ && \times\int d\zeta^{(-4)}_2 du_2 \Big\{ \square
 \d^{(2,2)}_A(1|2) - 4 f^2 \widetilde{Q}^+_1 G^{(1,1)}(1|2)
 Q^{+}_2\Big\}v^{++}_2\,,
 \label{S2V}
 \eea
and the hypermultiplet part
 \bea
 S_{2}^{Hyp}[V^{++}] = -\int d\zeta^{(-4)} du \, \tilde{h}^+
 \big(D^{++} + i V^{++}\big)h^+\,.
\label{S2H}
 \eea
We see that the quadratic part of the action  in the vector
multiplet sector $S^{Vect}_{2}$ is an analytic nonlocal functional
of quantum field $v^{++}$. It also contains an interaction between
background vector multiplet and  hypermultiplet through the
background-dependent Green function $G^{(1,1)}(V^{++})$.

The actions \eq{S2V} and \eq{S2H} specify the one-loop quantum
correction to the classical action \eq{S0}:
 \bea
 \Gamma^{(1)}[V^{++}, Q^+] = \f{i}{2} \Tr \ln \Big\{\square
 -4 f^2 \widetilde{Q}^+ G^{(1,1)}Q^+ \Big\} + i \Tr\ln \nb^{++}\,.
 \label{1loop}
 \eea
The expression \eq{1loop} is the starting point for studying the one-loop
effective action in the model \eq{S0}. In the next sections we will
calculate the divergent part of \eq{1loop}.

\section{Structure of one-loop counterterms}
In this section we analyze the superficial degree of divergence in
the model under consideration. The formal structure of Green
functions of the vector multiplet (\ref{VGreen}) and the
hypermultiplet (\ref{GREEN}) in $6D,\,{\cal N}=(1,0)$ gauge theory
is analogous to that in the four dimensional ${\cal N}=2$ case.
Hence, we can directly make use of the similar analysis in four
dimensional ${\cal N}=2$ theory \cite{BKO97}. As in the
four-dimensional theory, the Green functions in the case under
consideration contain enough number of Grassmann ${\delta}$-function
to prove the non-renormalization theorem according to which the loop
contribution to the supergraphs defining the effective action can be
written as a single integral over the total $6D, {\cal N}=(1,0)$
superspace.

Let us consider $L$-loop supergraph $G$ with $P$ propagators, $V$
vertices, $N_{Q}$ external hypermultiplet legs, and an arbitrary
number of gauge superfield external legs. We denote by $N_D$ the
number of spinor covariant derivatives acting on the external legs
as a result of integration by parts in the process of transforming
the contributions to a single integral over $d^8\theta$. The
superficial degree of divergence $\omega(G)$ of the supergraph $G$
can be found by counting the degrees of momenta in the loop
integrals.

The supergraph $G$ involves $L$ integrals over 6-momenta, which
contribute $6L$ to the degree of divergence. Each of the
hypermultiplet vertices contains one integration over $d^4\theta^+$.
Propagators of the gauge superfields contribute the factors $1/k^2$,
$(D^+)^4$, as well as the Grassmann $\delta$-functions. Similarly,
propagators of the hypermultiplet superfields contribute $1/k^2$,
$(D^+)^4$ for each of two harmonic arguments of propagator
(\ref{GREEN}) (eight $D^{+}$ - factors on a whole), and also the
Grassmann $\delta$-functions. From each hypermultiplet propagator we
take the operator $(D^+)^4$ and so complete $d^4\theta^+$ to
$d^8\theta$ in all hypermultiplets lines, except for the number
$\frac{1}{2}N_Q$ of them. Then we consider the corresponding
vertices and we take $\frac{1}{2}N_{Q}$ operators $(D^+)^4$ off the
propagators, which allow us to restore the integrations over
$d^8\theta$. After calculating the supergraph we will end up with a
single $d^8\theta$ integration. The other $V - 1$ integrations,
where $V$ is a total number of vertices, are done due to the
Grassmann $\delta$-functions. The remaining $P - V + 1 = L$
Grassmann $\delta$-functions survive. Each of them is killed by
eight $D^+_a$. Therefore, the number of remaining $D^+_a$ is $4 P -
2 N_{Q} - 8 L$. This implies that the superficial degree of
divergence is
\begin{equation}
\omega(G) = (6L - 2 P) + (2P - N_{Q} - 4L) -\frac{1}{2}N_D = 2L -
N_{Q} - \frac{1}{2}N_D,
\end{equation}
\noindent where $N_D$ is the number of the spinor covariant
derivatives acting on the external lines.

Equivalently, the degree of divergence can be calculated, using
dimension reasonings. Each gauge propagator brings $f^2,\, [f^2] =
m^{-2}$. The external gauge superfields are dimensionless,
$[V]=m^0$, while the dimension of hypermultiplets is $[q]= m^2$. The
effective action also contains a single integration over the full
superspace. Taking into account that $[d^6x] = m^{-6}$ and
$[d^8\theta] = m^4$ we see that
\begin{equation}
-\omega(G) =-2- 2P_V + 2N_{Q}+\frac{1}{2}N_D,
\end{equation}
where $P_V$ is the number of gauge propagators. For hypermultiplets
$N_{Q} = 2(-P_{Q} + V_{Q})$, so that
\begin{equation}\label{sdd}
\omega(G) = 2 - 2V + 2P - N_{Q} - \frac{1}{2}N_D = 2L - N_{Q} -
\frac{1}{2}N_D.
\end{equation}

Our aim is to calculate a divergent part of the one-loop effective
action, in this case the number of loops in Eq.\eq{sdd} is $L=1$.
Due to the analyticity of the hypermultiplet superfield, $D^+ q^+ =
0\,$, the number $N_D$ of spinor covariant derivatives acting on the
external legs is equal to zero, $N_D=0$. Thus, in our case the
superficial degree of divergence $\omega(G)$ \eq{sdd} is reduced to
 \bea
 \omega_{1-\rm loop}(G) = 2 - N_{Q} \,. \label{sdd2}
 \eea

Let us apply the relation (\ref{sdd2}) to the analysis of the one-loop
divergences. According to the general consideration  of ref. \cite{BIS}, the
possible contributions to divergent part of the effective action of
abelian theory is given by the following integral over the analytic
subspace of harmonic superspace:
 \bea
\Gamma_{div} = \int d\zeta^{(-4)}du\Big\{c_1 (F^{++})^2 + ic_2
\widetilde{Q}^+ F^{++}Q^+  + c_3 (\widetilde{Q}^+ Q^+)^2 \Big\}\,.
\label{div0}
 \eea
Here, the coefficients  $c_1, c_2, c_3$ depend on the regularization
parameters\footnote{In this paper we use the proper-time
regularization (see \cite{BP15}, \cite{BMP16} and references
therein) preserving the supersymmetry at least at one loop and are
interested in the logarithmic divergences only. One-loop logarithmic
divergences are known to be not susceptible to such subtleties of
quantum field theory as, e.g., presence of anomalies. We emphasize
that the regularization aspects of six dimensional theories deserve
a special attention, like those in four dimensional theories (see
e.g., discussion in \cite{BPS}). However, various choices of
regularization scheme do not affect the form of one-loop logarithmic
divergences which are the subject of our paper.}.

Let $N_{Q}=0$, then $\omega = 2$. The corresponding divergent
structure has to be quadratic in momenta and given by the full
${\cal N}=(1,0)$ superspace integral. The unique possibility is
 \bea
 \Gamma_1^{(1)} \sim \int d^{14}z \,du \, V^{--} \square V^{++}\,,
 \label{omega1}
 \eea
where $\square = \f12 (D^+)^4(D^{--})^2$. Integrating in \eq{omega1}
by parts, we can transfer the factor $(D^+)^4$ from d'Alembertian on
$V^{--}$ and use the definition of superfield $F^{++}$
\eqref{identity0}. Then we take one factor $D^{--}$ off the
second multiplier and make use of the zero-curvature condition
\eq{zeroc}. More precisely,
 \bea
 \Gamma_1^{(1)} &\sim&  \int d^{14}z \,du \,  F^{++} (D^{--})^2
 V^{++}  = - \int d^{14}z \,du \,  D^{--}F^{++} D^{--} V^{++}
 \nn \\
&& = -\int d^{14}z \,du \,  D^{--}F^{++} D^{++} V^{--} =
 \int d^{14}z \,du \,  D^{++} D^{--}F^{++}  V^{--}.
 \label{omega12}
 \eea
After that we commute the operators $D^{++}$ and $D^{--}$, use the
property $D^{++}F^{++}=0$ and obtain $ D^{++} D^{--}F^{++} = D_0
F^{++} = 2 F^{++}$. Finally, passing to the analytical subspace, we have
 \bea
 \Gamma_1^{(1)} &=& c_1 \int d\zeta^{(-4)} \,du \, (F^{++})^2.
 \label{omega13}
 \eea
The coefficient $c_1$ is divergent in the limit of removing the
regularization.

Let $N_{Q}=2$, then $\omega = 0$. The unique candidate divergent term involving no dependence on momenta and representable
as an integral over the full ${\cal N}=(1,0)$ superspace reads
 \bea
 \Gamma_2^{(1)} \sim \int d^{14}z \,du \, \widetilde{Q}^+ V^{--}Q^+\,.
 \label{omega2}
 \eea
Passing to the analytic subspace and using \eqref{identity0}, we
immediately obtain
 \bea
 \Gamma_2^{(1)} = i c_2 \int d\zeta^{(-4)} \,du \, \widetilde{Q}^+
 F^{++}Q^+\,,
 \label{omega22}
 \eea
where, once again, the coefficient $c_2$ is divergent in the limit
of removing the regularization. We see that the  contributions
\eq{omega13} and \eq{omega22} match with the general structure
\eq{div0} of the divergent part of the effective action.

For all other values of $N_{Q}$ the index $\omega$ is negative and
the corresponding Feynmann integrals are UV finite. In
particular, the divergent term of the form $(\widetilde{Q}^+Q^+)^2$ is
absent in the one-loop approximation. Such divergent terms could appear,
starting with two loops.

\section{Divergent part of the one-loop effective action}
In the previous section we discussed the general structure of the one-loop
contributions to the divergent part of effective action. Here we perform the direct calculation of
the coefficients $c_1$ and $c_2$ in (\ref{div0}).

The $(F^{++})^2$ part of the effective action depends only on the background
vector multiplet $V^{++}$ and is defined by the second term in eq.
\eq{1loop}. More precisely,
 \bea
\Gamma_{F^2}^{(1)}[V^{++}] &=& i\Tr\ln\nb^{++}
 =- i\Tr\ln G^{(1,1)}\,.
 \label{1loopV}
 \eea
Here $G^{(1,1)}$ is the superfield propagator for hypermultiplet
\eq{GREEN}. The details of calculation for \eq{1loopV} were discussed
in recent works \cite{BP15-1}, \cite{BMP16}. We consider an
arbitrary variation of the expression (\ref{1loopV})
 \bea
 \d \Gamma^{(1)}_{F^2}[V^{++}] &=& - i \, \Tr\d\, iV^{++}\, G^{(1,1)}
=  \int d\zeta_1^{(-4)} du_1\, \d V^{++}\, G^{(1,1)}
  (1|2)\Big|_{2=1}\,.
 \eea
Our aim is to calculate the divergent part of the effective action
\eq{1loopV}. In the proper-time regularization scheme \cite{BP15}, \cite{BMP16}, the divergences
are associated with the pole terms of the form
$\f{1}{\varepsilon}\,$, $\varepsilon \to 0$, where $\varepsilon =
6-d$ with space-time dimension $d$. Taking into account the
expression for Green function $G^{(1,1)}$ (\ref{GREEN}), one gets
 \be\label{div1}
\d\Gamma^{(1)}_{F^2}[V^{++}] = \int d\zeta_{1}^{(-4)} du_{1}\d
V^{++} \int_0^\infty
 d(is)(is\mu^2)^{\f\varepsilon2}
 e^{is\sB_1}(D_1^+)^4(D^+_2)^4\f{\delta^{14}(z_1-z_2)}{(u^+_1u^+_2)^3}\Big|^{2=1}_{\rm div}.
 \ee
Here $s$ is the proper-time parameter and $\mu$ is an arbitrary
regularization parameter of mass dimension. Like in the four- and
five-dimensional cases, one makes use of the identity (see \cite{KUZ} for details)
 \be\label{Delta}
(D^+_1)^4(D^+_2)^4\f{\d^{14}(z_1-z_2)}{(u^+_1u^+_2)^3}=(D^+_1)^4
\Big\{(u^+_1u^+_2)(\nb^-_1)^4 -(u^-_1u^+_2)\Omega_1^{--} -4\sB_1
\f{(u^-_1u^+_2)^2}{(u^+_1u^+_2)}\Big\} \d^{14}(z_1-z_2)\,,
 \ee
where we have introduced the notation
 \bea
 \Omega^{--} = i\nb^{ab}\nb^-_a\nb^-_b + 4W^{-a}\nb^-_a - (D^-_a
W^{- a})~. \label{O}
 \eea
One can show \cite{BM15} that only the first term in \eq{Delta} gives contribution to the divergent part of the one-loop effective action
 \bea \label{div2}
\d \Gamma^{(1)}_{F^2}[V^{++}] &=&  \int d\zeta_{1}^{(-4)}
du_{1}\d V^{++}(1) \nn \times \\
&& \times\int_0^\infty
 d(is)(is\mu^2)^{\f\varepsilon2}
 e^{is\sB_1}(u^+_1u^+_2)(D^+_1)^4 (D^-_1)^4 \delta^{14}(z_1-z_2)\Big|^{2=1}_{\rm div}.
  \eea
Those terms in the right hand side of (\ref{div2}) which produce the divergent part read
 \bea
e^{is\sB}(u^+_1u^+_2)e^{-is\sB}\Big|^{2=1}_{\rm div}  &=& -
i\f{(is)^2}{2}(\square F^{++}) -i\f{(is)^3}{6}\Big\{ 4 (\p^M\p^N
F^{++}) \p_M\p_N  \Big\}\,.
 \eea
Then we pass to  momentum representation of the delta function and
calculate the proper-time integral. This leads to the expression
 \bea
 \d \Gamma^{(1)}_{F^2}[V^{++}] = - \f{1}{3 (4\pi)^3 \varepsilon}\int d\zeta^{(-4)} du
\,\d V^{++}\, \square F^{++}\,. \label{div5}
 \eea

Let us compare \eq{div5} with \eq{div0}. Keeping in mind the
definition $F^{++} = (D^+)^4V^{--}\,$, we can transform the variation \eq{div5} to the form
 \bea
 \d \Gamma^{(1)}_{div} =  2 c_1\int d\zeta^{(-4)} du\, F^{++}
 (D^+)^4 \d V^{--}\,.
 \eea
Then we use the relation between $\d V^{--}$  and $\d V^{++}$
\eq{var} and the property $D^{++} F^{++}~=~0\,$. After that we restore the full $6D$,
$\cN=(1,0)$ superspace measure,
 \bea
  \d \Gamma^{(1)}_{div} =   c_1\int  dz du\,\, F^{++} (D^{--})^2
  \d V^{++} + \int du D^{++}(...)\,,
 \eea
and integrate by parts with respect to $(D^{--})^2$. Omitting the total
derivative terms and passing to the analytic subspace, we obtain
 \bea
  \d \Gamma^{(1)}_{ div} &=& c_1\int  dz du\,\, (D^{--})^2 F^{++}
  \d V^{++} \\
  &=& c_1 \int d\zeta^{(-4)} du\, (D^+)^4(D^{--})^2 F^{++}
  \d V^{++}.
 \eea
The derivatives $(D^+)^4$ act only on $(D^{--})^2 F^{++}$ because
$\d V^{++}$ is an analytic superfield. Then we use the definition of
analytic d'Alambertian $\square = \f12(D^+)^4 (D^{--})^2$ and
finally find
  \bea
  \d \Gamma^{(1)}_{div} &=& 2 c_1\int d\zeta^{(-4)} du\,
  \d V^{++} \, \square F^{++}\,.
  \label{div6}
 \eea
As is expected, the variation of the divergent part of effective
action \eq{div5} proved to have the same structure as \eq{div6}. Hence we
obtain, up to an unessential additive constant,
 \be
 \Gamma^{(1)}_{F^2}[V^{++}] =  -\f{1}{6 (4\pi)^3 \varepsilon}\int d\zeta^{(-4)} du
\, (F^{++})^2\,. \label{div7}
 \ee


The hypermultiplet-dependent part $\widetilde{Q}^+ F^{++}Q^+$ of the
one-loop counterterm arises from the first term in
\eq{1loop}\footnote{It is known that calculations of harmonic
supergraphs with hypermultiplet propagators require a certain care
related to coinciding harmonic singularities \cite{GNS}. As argued
in \cite{GNS} (see \cite{GIOS}, \cite{BIP} as well) this problem can
be avoided in all cases of interest. In our case we also do not face
such a problem.}. In order to calculate this contribution one
expands the logarithm in the first term \eq{1loop} up to the first
order and compute the functional trace
 \bea
\Gamma^{(1)}_{QFQ}[V^{++}, Q^+] &=& \f{i}{2} \Tr \ln \Big\{\square
 - 4f^2 \widetilde{Q}^+ G^{(1,1)}Q^+ \Big\} \nn \\
 &\approx& -2i f^2\int d\zeta^{(-4)} du\,  \widetilde{Q}^+Q^+
 \, \f{1}{\square}G^{(1,1)}(1|2)\Big|^{2=1}_{\rm div}\,.
  \eea
We again use the identity \eq{Delta} and consider only the first term
here, because  just this term is responsible for divergence:
 \bea
 \f{1}{\square}G^{(1,1)}(1|2)\Big|^{2=1}_{\rm div} &=&
  \f{1}{\square}
 \f{(D^+_1)^4(D^+_2)^4}{ \sB_1}\f{\d^{14}(z_1-z_2)}{(u^+_1u^+_2)^3}\Big|^{2=1}_{\rm
 div} \nn \\
 &=& \f{1}{\square}\f{(D^+_1)^4(D^-_1)^4}{\sB_1}(u^+_1u^+_2)
\delta^{14}(z_1-z_2)\Big|^{2=1}_{\rm div}\,.
 \eea
Now we observe that the first analytic d'Alembertian $\square$ in
the denominator comes from the pure vector multiplet part and does
not contain background fields. The second covariant d'Alembertian
$\sB$ in the denominator emerges from the Green function for
hypermultiplet after non-local change of variables \eq{replac}. This
operator depends on the background vector multiplet as in \eqref{Box1}.

To calculate the divergent part of the expression under
consideration it suffices to take into account only two first
terms in the $\sB$, namely
$$\sB = \square + i F^{++} \nb^{--}+\ldots.$$ Other two terms do not
contribute to the divergent part of one-loop effective action in the
point-coincidence limit. We expand the operator $\f{1}{\sB}$ up to
the first order in $iF^{++}\nb^{--}$ and act by it on
the harmonic distribution $(u^+_1u^+_2)$. Using  properties of Grassmann
delta-function, $(D^+_1)^4 (D^-_1)^4
\delta^{8}(\theta_1-\theta_2)\Big|_{2=1} = 1$, we obtain
 \bea
 \f{(D^+_1)^4(D^-_1)^4}{\square(\square + i F^{++}\nb^{--}+..)}(u^+_1u^+_2)
\delta^{14}(z_1-z_2)\Big|^{2=1}_{\rm div}
 = - i F^{++} \f{(u^-_1u^+_2)}{\square^3} \d^6(x_1-x_2)\Big|_{2=1} \,.
 \eea
Then one uses the momentum representation of the space-time
$\delta$-function and calculates the momentum integral in
the $\varepsilon$-regularization scheme. It leads to
 \be\f{1}{\square^3}
 \d^6(x_1-x_2)\Big|_{2=1} = \f{i}{(4\pi)^3}\f{1}{\varepsilon}\,, \quad
 \varepsilon \to 0\,.
  \ee
  As a result, one gets
 \bea
\Gamma^{(1)}_{QFQ}[V^{++}, Q^+] &=&\f{ 2 if^2}{(4\pi)^3 \varepsilon}
\int d\zeta^{(-4)}du\, {\widetilde Q}^+ F^{++} Q^+\,. \label{div8}
 \eea

Summing up the contributions \eq{div7} and \eq{div8}, we finally
obtain
 \bea
 \Gamma^{(1)}_{div}[V^{++}, Q^+] = -\f{1}{ 6 (4\pi)^3 \varepsilon} \int d\zeta^{(-4)}du\,
 \Big\{(F^{++})^2 - 12\, if^2 {\widetilde Q}^+ F^{++} Q^+ \Big\}.
 \eea
If the background hypermultiplet vanishes, the divergent part of the
effective action is proportional to the classical equation of motion
$F^{++}=0$. Therefore the divergence as a whole can be eliminated by
a field redefinition ($\delta{V^{++}} \sim
\frac{1}{\varepsilon}F^{++})$ in the classical action and the theory
under consideration is one-loop finite on shell, in accordance with
the results of ref. \cite{HS}. However, if the background
hypermultiplet does not vanish, we obtain, after some field
redefinition proportional to the equation of motion, the divergent
part of on-shell effective action in the form $\Gamma^{(1)}_{div}
\sim \frac{1}{\varepsilon}\int d\zeta^{(-4)}du ({\widetilde
Q}^{+}Q^{+})^2$. Thus, the on-shell divergence in the hypermultiplet
sector cannot be eliminated and the full theory is not finite even
at the one-loop level.

\section{Summary and outlook}
Let us briefly summarize the results obtained. We have considered
the six-dimensional ${\cal N}=(1,0)$ supersymmetric theory of the
abelian  vector multiplet coupled to hypermultiplet in the $6D,
\,{\cal N}=(1,0)$ harmonic superspace formulation. We have studied
the quantum effective action involving dependence on both the vector
multiplet and the hypermultiplet superfields. The corresponding
background field method in harmonic superspace was formulated, such
that it allows one to preserve manifest gauge invariance and
supersymmetry at all stages of calculating the effective action. It
is important to point out that the superfield propagators in the
theory under consideration have, in the sector of anticommuting
variables and harmonics, the same structure as the propagators in
$4D,\, {\cal N}=2$ SYM theory. It leads to $6D, \,{\cal N}=(1,0)$
renormalization theorem, which states that the contribution of any
supergraph in the theory under consideration can be written as a
single integral over anticommuting variables of the full $6D,\,
{\cal N}=(1,0)$ superspace. Using this result, we have calculated
the superficial degree of divergences and analyzed the structure of
one-loop counterterms in both the vector multiplet and the
hypermultipelt sectors. It was shown, in particular, that one of the
possible divergent counterterms in the purely hypermultiplet sector,
which is allowed on the supersymmetry and dimension grounds
\cite{BIS}, is actually prohibited at one loop.

We have developed an efficient manifestly gauge invariant and
${\cal N}=(1,0)$ supersymmetric technique to calculate the one-loop
effective action. As an application of this technique, we found the
one-loop divergences of the theory under consideration. The results
completely match the analysis of the general structure of divergences based
on considering superficial degree of divergences. It was shown that,
if the background hypermultiplet superfield does not vanish, the one-loop
divergences cannot be eliminated by any field redefinition and the
theory is not one-loop finite.

Let us discuss some possible generalizations and extensions of the
results obtained. As the next step, it is quite natural to study the
structure of the effective action for the non-abelian $6D, \,{\cal
N}=(1,0)$ SYM theories. All such theories admit a formulation in
$6D, \,{\cal N}=(1,0)$ harmonic superspace. The background field
method can be developed quite analogously to the abelian case and
the one-loop divergences can be calculated. The basic difference
from the abelian theory considered here will be a self-interaction
of vector multiplet and a non-trivial ghost contribution to the
effective action, which can change the relative coefficient between
the $(F^{++})^2$ and the ${\widetilde Q}^+ F^{++} Q^+$ terms in the
one-loop divergent part. We expect that the purely hypermultiplet
contribution to the divergent part of the one-loop effective action
in non-abelian theory will be absent as in the abelian theory.
Besides the divergent part of effective action, it would be
interesting to study the finite contributions to low-energy
effective action, which have never been considered before.

It would be extremely interesting to study the effective action in
$6D,\, {\cal N}=(1,1)$ SYM theory. Such a theory can be formulated
in $6D, \,{\cal N}=(1,0)$ harmonic superspace in terms of ${\cal
N}=(1,0)$ analytic harmonic superfields, viz. the gauge connection
$V^{++}$ and the hypermultiplet $q^+, \tilde{q}^+$, both in the
adjoint representation \cite{BIS}. This theory exhibits the manifest
off-shell ${\cal N}=(1,0)$ supersymmetry and an additional hidden
on-shell ${\cal N}=(0,1)$ supersymmetry, and in many aspects is
analogous to $4D, \,{\cal N}=4$ SYM theory \cite{GIOS}. It was
shown, based solely upon the invariance of the effective action
under both manifest and hidden supersymmetries, that ${\cal
N}=(1,1)$ SYM theory is one-loop finite. It would be tempting to
analyze the divergences of ${\cal N}=(1,1)$ SYM theory within the
quantum setting and explicitly calculate the one-loop counterterms
(in parallel with constructing the full quantum ${\cal N}=(1,1)$ SYM
effective action).

It is well known that the $6D, \,{\cal N}=(1,0)$ supersymmetric theories are
anomalous (see discussions of chiral anomalies in higher dimensional
supersymmetric theories in refs. \cite{FK}). It would be interesting
to study such anomalies in the harmonic superspace formulation of $6D, {\cal N}=(1,0)$ SYM coupled to hypermultiplets
and show, by a direct quantum field
theoretical analysis, that the ${\cal N}=(1,1)$ SYM theory is anomaly-free.
We are going to tackle all these problems in the forthcoming works.


\section*{Acknowledgements}
\noindent I.L.B and E.A.I are grateful to Kelly Stelle for useful
comments and discussion. They also thank the organizers of the MIAPP
program "Higher spin theory and duality" (May 2-27, 2016) for the
hospitality in Munich at the early stages of this work. The authors
acknowledge support from the Russian Science Foundation grant,
project No 16-12-10306.


\end{document}